\documentclass{emulateapj}
\usepackage{apjfonts}
\usepackage{rotating}
\usepackage{multirow}

\lefthead{PARK ET AL.}
\righthead{AMBIGUITY IN PLANETARY MICROLENSING INTERPRETATIONS}

\begin{document}
\title{OGLE-2012-BLG-0455/MOA-2012-BLG-206: MICROLENSING EVENT WITH AMBIGUITY 
IN PLANETARY INTERPRETATIONS CAUSED BY INCOMPLETE COVERAGE OF PLANETARY SIGNAL}

\author{
H. Park$^{U1}$, 
C. Han$^{U1,U,\dag}$, 
A. Gould$^{U2,U}$, 
A. Udalski$^{O1,O}$, 
T. Sumi$^{M1,M}$, 
P. Fouqu\'e$^{P1}$, \\
and\\
J.-Y. Choi$^{U1}$, 
G. Christie$^{U3}$, 
D. L. Depoy$^{U4}$, 
Subo Dong$^{U5}$, 
B. S. Gaudi$^{U2}$,
K.-H. Hwang$^{U1}$, 
Y. K. Jung$^{U1}$, 
A. Kavka$^{U2}$, \\
C.-U. Lee$^{U6}$,
L. A. G. Monard$^{U7}$, 
T. Natusch$^{U3,U8}$, 
H. Ngan$^{U3}$, 
R. W. Pogge$^{U2}$, 
I.-G. Shin$^{U1}$, 
J. C. Yee$^{U2,U9,\clubsuit}$, \\
(The $\mu$FUN Collaboration)\\
M. K. Szyma\'nski$^{O1}$, 
M. Kubiak$^{O1}$, 
I. Soszy\'nski$^{O1}$, 
G. Pietrzy\'nski$^{O1,O2}$, 
R. Poleski$^{O1,U2}$, 
K. Ulaczyk$^{O1}$, 
P. Pietrukowicz$^{O1}$,
S. Koz{\l}owski$^{O1}$, 
J. Skowron$^{O1}$, 
{\L}. Wyrzykowski$^{O1,O3}$, \\
(The OGLE Collaboration)\\
F. Abe$^{M2}$,
D. P. Bennett$^{M3}$,
I. A. Bond$^{M7}$,
C. S. Botzler$^{M4}$,
P. Chote$^{M5}$,
M. Freeman$^{M4}$,
A. Fukui$^{M6}$,
D. Fukunaga$^{M2}$, \\
P. Harris$^{M5}$,
Y. Itow$^{M2}$,
N. Koshimoto$^{M1}$,
C. H. Ling$^{M7}$,
K. Masuda$^{M2}$,
Y. Matsubara$^{M2}$,
Y. Muraki$^{M2}$,
S. Namba$^{M1}$, \\
K. Ohnishi$^{M8}$, 
N. J. Rattenbury$^{M4}$,
To. Saito$^{M9}$,
D. J. Sullivan$^{M5}$,
W. L. Sweatman$^{M7}$,
D. Suzuki$^{M1}$,
P. J. Tristram$^{M10}$, \\
K. Wada$^{M1}$, 
N. Yamai$^{M11}$,
P. C. M. Yock$^{M4}$,
A. Yonehara$^{M11}$ \\
(The MOA Collaboration)\\
}

\bigskip\bigskip
\affil{$^{U1}$Department of Physics, Institute for Astrophysics, Chungbuk National University, Cheongju 371-763, Korea}
\affil{$^{U2}$Department of Astronomy, The Ohio State University, 140 West 18th Avenue, Columbus, OH 43210, USA}
\affil{$^{O1}$Warsaw University Observatory, Al. Ujazdowskie 4, 00-478 Warszawa, Poland}
\affil{$^{M1}$Department of Earth and Space Science, Osaka University, Osaka 560-0043, Japan}
\affil{$^{P1}$IRAP, CNRS, Universit\'e de Toulouse, 31400 Toulouse, France}
\affil{$^{U3}$Auckland Observatory, Auckland, New Zealand}
\affil{$^{U4}$Department of Physics and Astronomy, Texas A\&M University, College Station, TX 77843, USA}
\affil{$^{U5}$Kavli Institute for Astronomy and Astrophysics, Peking University, Yi He Yuan Road 5, Hai Dian District, Beijing 100871, China}
\affil{$^{U6}$Korea Astronomy and Space Science Institute, 776 Daedukdae-ro, Yuseong-gu, Daejeon 305-348, Korea}
\affil{$^{U7}$Kleinkaroo Observatory, Calitzdorp, and Bronberg Observatory, Pretoria, South Africa}
\affil{$^{U8}$Institute for Radiophysics and Space Research, AUT University, Auckland, New Zealand}
\affil{$^{U9}$Harvard-Smithsonian Center for Astrophysics, 60 Garden St., Cambridge, MA 02138, USA}
\affil{$^{O2}$Universidad de Concepci\'on, Departamento de Astronomia, Casilla 160-C, Concepci\'{o}n, Chile}
\affil{$^{O3}$Institute of Astronomy, University of Cambridge, Madingley Road, Cambridge CB3 0HA, UK}
\affil{$^{M2}$Solar-Terrestrial Environment Laboratory, Nagoya University, Nagoya, 464-8601, Japan}
\affil{$^{M3}$Department of Physics, University of Notre Dame, 225 Nieuwland Science Hall, Notre Dame, IN 46556-5670, USA}
\affil{$^{M4}$Department of Physics, University of Auckland, Private Bag 92-019, Auckland 1001, New Zealand}
\affil{$^{M5}$School of Chemical and Physical Sciences, Victoria University, Wellington, New Zealand}
\affil{$^{M6}$Okayama Astrophysical Observatory, National Astronomical Observatory of Japan, Asakuchi, Okayama 719-0232, Japan}
\affil{$^{M7}$Institute of Information and Mathematical Sciences, Massey University, Private Bag 102-904, North Shore Mail Centre, Auckland, New Zealand}
\affil{$^{M8}$Nagano National College of Technology, Nagano 381-8550, Japan}
\affil{$^{M9}$Tokyo Metropolitan College of Aeronautics, Tokyo 116-8523, Japan}
\affil{$^{M10}$Mt. John University Observatory, P.O. Box 56, Lake Tekapo 8770, New Zealand}
\affil{$^{M11}$Department of Physics, Faculty of Science, Kyoto Sangyo University, 603-8555, Kyoto, Japan}

\affil{$^{U}$The $\mu$FUN Collaboration}
\affil{$^{O}$The OGLE Collaboration}
\affil{$^{M}$The MOA Collaboration}
\affil{$^{\clubsuit}$Sagan Fellow}
\affil{$^{\dag}$Corresponding author}

\begin{abstract}
Characterizing a microlensing planet is done from modeling an observed lensing light 
curve. In this process, it is often confronted that solutions of different lensing parameters 
result in similar light curves, causing difficulties in uniquely interpreting the lens 
system, and thus understanding the causes of different types of degeneracy is important. In 
this work, we show that incomplete coverage of a planetary perturbation can result 
in degenerate solutions even for events where the planetary signal is detected with a 
high level of statistical significance. We demonstrate the degeneracy for an actually 
observed event OGLE-2012-BLG-0455/MOA-2012-BLG-206. The peak of this high-magnification 
event $(A_{\rm max}\sim400)$ exhibits very strong deviation from a point-lens model 
with $\Delta\chi^{2}\gtrsim4000$ for data sets with a total number of measurement 6963. 
From detailed modeling of the light curve, we find that the deviation can be explained 
by four distinct solutions, i.e., two very different sets of solutions, each with a two-fold 
degeneracy. While the two-fold (so-called ``close/wide'') degeneracy is well-understood, 
the degeneracy between the radically different solutions is not previously known. The model 
light curves of this degeneracy differ substantially in the parts that were not covered by 
observation, indicating that the degeneracy is caused by the incomplete coverage of the 
perturbation. It is expected that the frequency of the degeneracy introduced in this work 
will be greatly reduced with the improvement of the current lensing survey and follow-up 
experiments and the advent of new surveys.
\end{abstract}


\section{INTRODUCTION}

Gravitational microlensing is one of important methods to detect and characterize 
extrasolar planets. Due to its sensitivity to planets that are otherwise difficult  
to detect, the microlensing method is complementary to other methods. In particular, 
the method is sensitive to planets of low-mass stars located at or beyond the snow line, 
low-mass planets including terrestrial planets \citep{jung14}, and even free-floating 
planets \citep{sumi11}. For general review of planetary microlensing, see \citet{gaudi12}.

The microlensing signal of a planet is usually a short-term perturbation to the smooth and 
symmetric standard light curve of the primary-induced lensing event \citep{mao91,gould92b}. 
The planetary perturbation occurs when the source approaches planet-induced caustics, 
that represent the positions on the source plane at which the magnification of a point 
source would become infinite. For a lens composed of a star and a planet, caustics 
form a single or multiple sets of closed curves each of which is composed of concave 
curves that meet at cusps. The number, size, and shape of caustics vary depending on 
the separation and the mass ratio between the planet and its host star. For a given 
planetary system, planetary perturbations further vary depending on how the source 
approaches the lens. As a result, planets exhibit very diverse signals in lensing 
light curves.

Due to the immense diversity of planetary signals, characterizing a microlensing planet is 
a difficult task. This characterization is done from modeling in which an observed lensing 
light curve is compared to numerous theoretical curves resulting from various combinations 
of the parameters describing the lens and the source. In this process, it is often confronted 
that solutions of different lensing parameters result in similar light curves and can explain 
the observed light curve. This degeneracy problem causes difficulty in the unique interpretation 
of the lens system. Therefore, understanding the causes of various types of degeneracy is 
very important.

Up to now, it is known that there exist three broad categories of degeneracy in the interpretation 
of planetary microlensing signals. The first category corresponds to the case for which the degeneracy 
occurs when different planetary systems induce similar caustics. Good examples are the ``close/wide'' 
degeneracy for binary-lens events \citep{griest98,dominik99,an05} and the ``ecliptic'' degeneracy 
for events affected by parallax effects \citep{skowron11}. The second category is that the degeneracy 
occurs when light curves accidentally appear to be similar despite the fact that the caustics of the 
degenerate solutions are very different. \citet{choi12} presented two examples of events for which 
an observed perturbation could be interpreted by either a planetary or a binary companion. The third 
category is that perturbations can be interpreted by solutions of totally different origins. 
A good example is the binary-lens/binary-source degeneracy \citep{gaudi98,gaudi04,hwang13}.

In this work, we show that incomplete coverage of a perturbation can also result in 
degenerate solutions even for events where the planetary signal is detected with a high 
level of statistical significance. We demonstrate the degeneracy for an actually observed 
event OGLE-2012-BLG-0455/MOA-2012-BLG-206.

\section{Observation}

The microlensing event OGLE-2012-BLG-0455/MOA-2012-BLG-206 occurred on a star located 
close to the Galactic center with equatorial coordinates  $(\alpha,\delta)_{\rm J2000}
=(17^{\rm h}51^{\rm m}32^{\rm s}\hskip-2pt.42, -28^{\circ}33'42''\hskip-2pt.3)$, 
corresponding to the Galactic coordinates $(l,b)=(0.99^\circ, -0.92^\circ)$. 
The lensing induced brightening of the source star was first noticed on April 16, 2012 
from the lensing survey conducted by the Optical Gravitational Lensing Experiment 
\citep[OGLE:][]{udalski03} using the 1.3m Warsaw telescope of Las Campanas Observatory in Chile. 
The event was independently detected from the survey done by the Microlensing Observations 
in Astrophysics \citep[MOA:][]{bond01,sumi03} group using the 1.8m telescope of Mt. John 
Observatory in New Zealand. Based on real-time modeling of OGLE and MOA data (posted on 
their web sites\footnotemark[1]$^,$\footnotemark[2]), 
the Microlensing Follow-Up Network \citep[$\mu$FUN:][]{gould06} issued a second level alert just 9 
hours before the peak, predicting that the event would be very high magnification ($A_{\rm max}>300$) 
and so would be extremely sensitive to planets \citep{griest98}. In response to the high-magnification 
alert, data were taken first by using the 0.36m telescope of Kleinkaroo Observatory (KKO) in South Africa 
and subsequently by using the 1.3m SMARTS telescope of Cerro Tololo Inter-American Observatory (CTIO) 
in Chile and the 0.4m telescope of Auckland observatory in New Zealand. From follow-up observations, 
the peak of the event was densely covered, especially by the CTIO data, that are composed of 55 images 
in {\it I}, 8 images in {\it V}, and 295 images in {\it H} band. The total number of measurement is 6963. 
However, the coverage is not complete because the event occurred during the early Bulge season when the 
duration of Galactic-bulge visibility was short and follow-up observation in other parts of the Earth was 
not fully operational.
\footnotetext[1]{
http://ogle.astrouw.edu.pl}
\footnotetext[2]{
http://www.phys.canterbury.ac.nz/moa} 

\begin{figure}[ht]
\epsscale{1.15}
\plotone{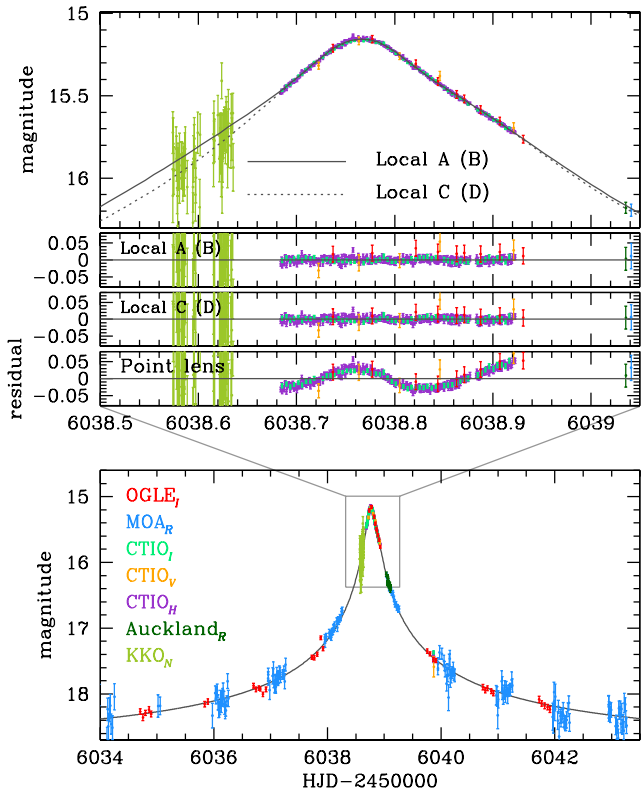}
\caption{\label{fig:one}
Light curve of OGLE-2012-BLG-0455/MOA-2012-BLG-206. In the legends 
indicating observatories, the subscript of each observatory denotes 
the passband. The top panels show the zoom of the peak region where 
solid and dotted curves represent the models of the degenerate solutions. 
The three middle panels show the residuals from the two degenerate 
planetary solutions and the point-lens model. 
}\end{figure}

The OGLE and MOA data were reduced using photometry codes developed by the individual groups, 
based on the Difference Image Analysis technique \citep{alard98,wozniak00,bond01}. The $\mu$FUN 
data were initially reduced by DoPHOT pipeline \citep{schechter93} and were reprocessed using 
the pySIS package \citep{albrow09} to refine the photometry. Photometric errors estimated by 
different photometry systems may vary. Futhermore, error bars of each data set may deviate from 
the dispersion of the data points due to systematics in the photometry system. In order to use 
data sets collected from different observatories, we therefore normalize error bars. For 
this, we first add a quadratic term so that the cumulative distribution of $\chi^{2}$ ordered 
by magnification is approximately linear to ensure that the dispersion of the data points is 
consistent with error bars regardless of the source brightness. We then rescale the errors so 
that $\chi^{2}$ per degree of freedom ($\chi^{2}$/dof) for each data set becomes unity to ensure 
that each data set is fairly weighted according to error bars.

\footnotetext[3]{
We note that the apparent lensing magnification $A_{\rm obs}\sim35$, 
corresponding to the magnitude change $\sim3.5$ mag, is much smaller than the measured magnification 
$A_{\rm max}\sim400$, because the lensed star is heavily blended with other neighboring stars.
}

In Figure~\ref{fig:one}, we present the light curve of the event. The event reached a magnification 
$A_{\rm max}\sim400$ at the peak.\footnotemark[3] At a glance, the light curve appears to have a 
standard form of an event caused by a point mass. However, a single-lens fit leaves significant 
residual near the peak. Such a deviation at the peak is typically produced by either a planetary 
or a binary companion to the primary lens.

\section{Modeling}

Keeping the possible cause of the perturbation in mind, we analyze the light curve 
based on two point-mass lens modeling. Basic description of a binary-lens event requires 
7 parameters. Among them, the first three describe the lens-source approach, including 
the time of the closest approach of the source to a reference position of the binary lens, 
$t_{0}$, the lens-source separation at $t_{0}$ in units of the angular Einstein radius 
$\theta_{\rm E}$ of the lens, $u_{0}$, and the time required for the source to cross the 
Einstein radius, $t_{\rm E}$ (Einstein time scale). Another three parameters describe the 
two-point lens, including the projected binary separation in units of the Einstein radius, 
$s$, the mass ratio between the lens components, $q$, and the angle between the source 
trajectory and the binary axis, $\alpha$. The last parameter is the ratio of the angular 
source radius $\theta_{\ast}$ to the Einstein radius, $\rho_{\ast}=\theta_{\ast}/\theta_{\rm E}$ 
(normalized source radius), which is needed to describes the effect of the extended source 
on the light curve.

\begin{figure}[ht]
\epsscale{1.15}
\plotone{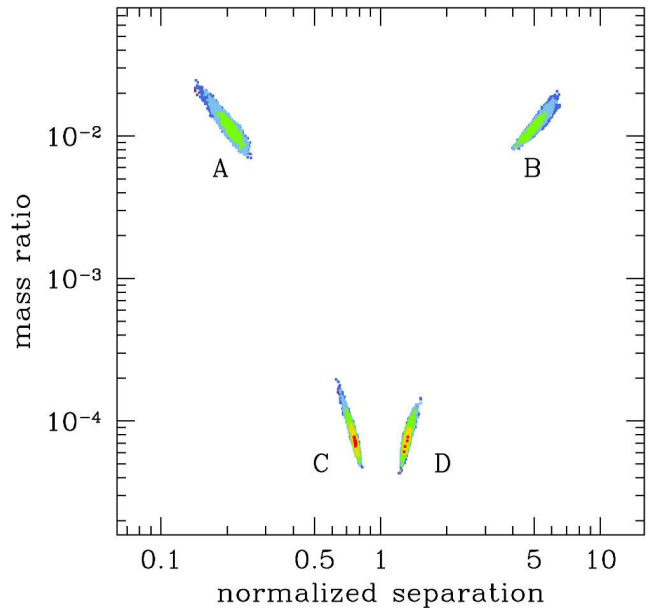}
\caption{\label{fig:two}
Distribution of $\Delta\chi^{2}$ in the parameter space of the 
normalized projected separation $s$ and the mass ratio $q$. 
Different contours correspond to $\Delta\chi^{2}<1$ (red), 4 (yelloew), 
9 (green), 16 (light blue), 25 (blue), and  36 (purple), respectively.
}\end{figure}

Besides the basic lensing parameters, additional parameters are often needed to describe subtle 
deviations of lensing light curves caused by second-order effects. One such an effect is 
the orbital motion of a binary lens, which induces variation of the caustic shape during 
the magnification phase \citep{albrow00,an05,penny11,shin11,park13}. Another effect is caused 
by the orbital motion of the Earth which results in deviation of the source motion from 
rectilinear \citep{gould92a}. The latter effect is often referred to as parallax effect. 
We find that these effects are not important for OGLE-2012-BLG-0455/MOA-2012-BLG-206 mainly 
due to the relatively short duration of the event and moderate photometric quality in the 
wing and baseline.

The search for the best-fit solution of the lensing parameter is conducted in two steps. 
In the first step, we conduct grid search in the $(s,q,\alpha)$ parameter space in order to 
locate all possible local minima. In this process, the remaining parameters $(t_{0}, u_{0}, 
t_{\rm E}, \rho_{\ast})$ are searched by a downhill approach to yield minimum $\chi^{2}$ 
at each grid point. In the second step, we investigate the individual local minima found from the 
initial search. At this stage, we refine each local minimum by allowing all parameters to 
vary. For $\chi^{2}$ minimization, we use the Markov Chain Monte Carlo (MCMC) method.

\begin{deluxetable*}{lrrrr}
\tablecaption{Lensing parameters of 4 degenerate solutions \label{table:one}}
\tablewidth{0pt}
\tablehead{
\multicolumn{1}{c}{Parameters} &
\multicolumn{1}{c}{Local A}  &
\multicolumn{1}{c}{Local B} &
\multicolumn{1}{c}{Local C} & 
\multicolumn{1}{c}{Local D}  
}
\startdata
$\chi^2/{\rm dof}$         & 6963.6/6956            & 6962.7/6956            & 6957.6/6956            & 6957.7/6956           \\
$t_{0}$ (HJD-2450000)      & 6038.7683 $\pm$ 0.0004 & 6038.7689 $\pm$ 0.0004 & 6038.7768 $\pm$ 0.0005 & 6038.7770 $\pm$ 0.0005\\
$u_{0}$ ($10^{-3}$)        &    2.32 $\pm$ 0.17     &    2.21 $\pm$ 0.10     &    2.14 $\pm$ 0.12     &    2.23 $\pm$ 0.17    \\
$t_{\rm E}$ (days)         &    47.4 $\pm$ 3.3      &    50.3 $\pm$ 2.3      &    50.1 $\pm$ 2.6      &    48.0 $\pm$ 3.6     \\
$s$                        &    0.23 $\pm$ 0.02     &    4.99 $\pm$ 0.39     &    0.77 $\pm$ 0.02     &    1.33 $\pm$ 0.04    \\
$q$ ($10^{-3}$)            &    9.55 $\pm$ 2.26     &   11.30 $\pm$ 1.74     &    0.07 $\pm$ 0.01     &    0.08 $\pm$ 0.01    \\
$\alpha$ (rad)             &   4.777 $\pm$ 0.009    &   4.782 $\pm$ 0.009    &   4.209 $\pm$ 0.003    &   4.211 $\pm$ 0.004   \\
$\rho_{\ast}$ ($10^{-3}$)  &    1.56 $\pm$ 0.12     &    1.35 $\pm$ 0.09     &    1.08 $\pm$ 0.06     &    1.11 $\pm$ 0.09    
\enddata                                              
\end{deluxetable*}

\begin{figure}[ht]
\epsscale{1.15}
\plotone{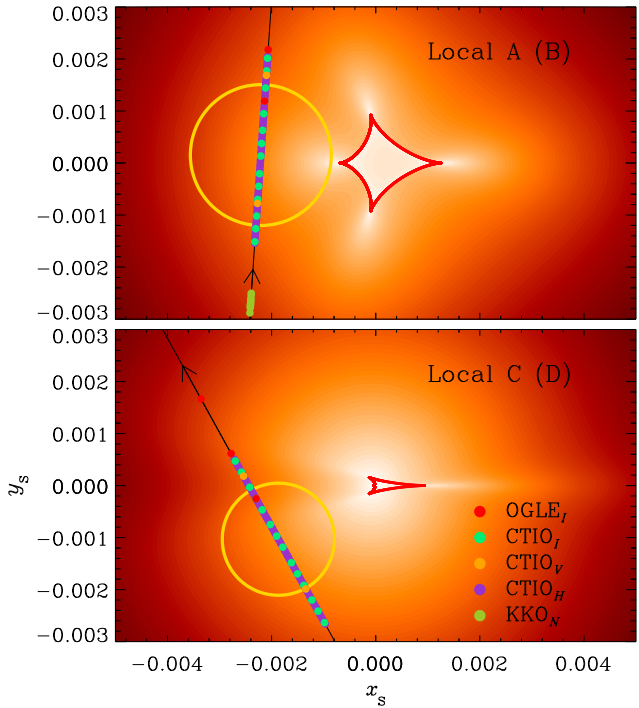}
\caption{\label{fig:three}
Geometries and magnification patterns for the best-fit models of Local A 
or B (upper panel), and C or D (lower panel). The brighter tone denotes higher 
magnifications. The closed red curves represent the caustics. In each panel,
the straight line with an arrow represents the source trajectory. The size of the 
empty yellow circle represents the source size. The dots on the trajectories 
represent the source positions at the times when data were taken. 
}\end{figure}

We compute finite-source magnifications by using the inverse ray-shooting method. In this numerical 
method, uniform rays are shot from the image plane, bent by the lens equation, and arrive 
at the source plane. We note that the term ``inverse'' is used to denote that rays are traced 
{\it backward} from the image plane to the source plane. Then, the magnification affected by 
the extended source is computed as the ratio between the number densities of rays on the source 
surface and on the image plane \citep{schneider86,kayser86,wambsganss97}. The lens equation of 
a binary lens is expressed as
\begin{equation}
\zeta=z-\sum\limits_{i=1}^2 {\epsilon_{i} \over \overline{z}-\overline{z}_{{\rm L},i}},
\label{eq1}
\end{equation}
where $\zeta$, $z_{{\rm L},i}$, and $z$ represent the complex notations of the source, lens, 
and image positions, the overbar denotes complex conjugate, $\epsilon_{i}$ are the mass fractions 
of the lens components, and the index $i=1,2$ denote the individual lens components. In computing 
finite magnifications, we consider the limb-darkening effect of the source star by modeling 
the surface brightness profile as
\begin{equation}
S_{\lambda}\propto 1-\Gamma_{\lambda}\left( 1-{3\over2}\cos\psi \right),
\label{eq2}
\end{equation}
where $\Gamma_{\lambda}$ is the linear limb-darkening coefficient, $\lambda$ is the passband, 
and $\psi$ is the angle between the line of sight toward the source star and the normal to 
the source surface. 
The limb-darkening coefficients are adopted from \citet{claret00} considering 
the source type estimated based on the location in the color-magnitude diagram. It is estimated 
that the source type is an F-type main-sequence star with the dereddened color $(V-I)_{0}=0.60$ and 
{\it I} magnitude $I_{0}=18.6$. Based on the source type, we adopt the coefficients 
$\Gamma_{V}=0.497$, $\Gamma_{R}=0.421$, $\Gamma_{I}=0.347$, and $\Gamma_{H}=0.199$. For the MOA 
data, which used a non-standard filter system, we choose a mean value of the {\it R} and {\it I} 
band coefficients, i.e., $(\Gamma_{R}+\Gamma_{I})/2$.

\section{Results}

From detailed analysis of the light curve, we find that the light curve significantly 
deviates from a standard point-mass model with $\Delta\chi^{2}\gtrsim4000$. However, despite 
such a strong signal, interpreting the deviation is difficult due to the existence of very 
degenerate local minima in the parameter space. Figure~\ref{fig:two} shows the local 
minima presented as $\Delta\chi^{2}$ distribution in the $(s,q)$ parameter space. 
It is found that there exist 4 distinct local minima. We mark the individual minima as 
A, B, C, and D. In Table~\ref{table:one}, we list the lensing parameters of the individual 
local minima. We note that the mass ratios of all the minima are less than $10^{-2}$, 
implying that the companion is in the planetary mass regime. However, the degeneracy 
among the local solutions is very severe with $\Delta\chi^{2}\lesssim5$ for dof=6956\footnotemark[4] 
and thus the characteristics of the planet cannot be uniquely determined.
\footnotetext[4]{
We note that the source star of the event had been observed since 2010 and 
thus there exist much more data points not shown in Figure~\ref{fig:one}.
}

Among the local solutions, the degeneracy between the A-B and C-D pairs are already known. 
For each of these pairs, the mass ratios are similar but the projected separations have 
opposites signs of $\log s$, i.e. $s\leftrightarrow s^{-1}$. For such pairs of binary lenses, 
the caustics located near the primary lens induced by the close $(s<1)$ and wide $(s>1)$ 
planetary companions are similar both in size and shape, causing degeneracy in the resulting 
light curve. This degeneracy, known as the close/wide degeneracy, is caused by the invariance 
of the caustic under the $s\leftrightarrow s^{-1}$ transformation \citep{griest98,dominik99,an05,chung05}.

On the other hand, the degeneracy between the A-C and B-D pairs are not previously known. 
For each of these pairs, the lens systems of the individual local solutions have widely 
different characteristics. For example, the values of the separation and mass ratio are 
$(s,q)=(0.23, 9.50\times10^{-3})$ for the local solution ``A'' while the values are 
$(s,q)=(0.77, 0.07\times10^{-3})$ for the solution ``C''. Based on the mass ratio, the 
individual solutions imply that the planet is either a super Jupiter or a Neptune-mass 
planet if the primary is a normal star. Due to the wide difference in the planet parameters, 
the caustics and the magnification patterns around the caustics of the degenerate 
solutions are greatly different as shown in Figure~\ref{fig:three}. Despite the difference 
in caustics, we find that the two solutions are very degenerate with $\Delta\chi^{2}\lesssim5$.

Although the former degeneracy is severe because it is intrinsically rooted in the lens equation, 
the latter degeneracy results from widely different lens systems and thus it might be that the 
degeneracy could be resolved with additional information. We, therefore, conduct three additional 
tests to check the feasibility of resolving the degeneracy.

The first test is to compare limb-darkening effects of the source star. For the high mass-ratio 
solutions (local A and B), the source approaches the caustic close enough for the edge of the source 
star almost to touch the caustic. For the low mass-ratio solutions (local C and D), on the other hand, 
the source-caustic separation is relatively wide. Then, the limb-darkening effect would be more 
important for the high mass-ratio solution than the low mass-ratio solution. We investigate the 
limb-darkening effect by measuring the color variation in the CTIO {\it I} and {\it H} data taken 
during the caustic approach. Unfortunately, the expected color variation from the models is 
substantially smaller than the photometric errors. Therefore, this method cannot be applied 
to resolve the degeneracy.

The second test is to compare source fluxes estimated from the degenerate solutions. If they 
are different, high resolution imaging from either space-based or ground-based adaptive optics 
observation would enable one to distinguish the solutions by resolving blended stars. However, 
we find that the source and blend fluxes for the 2 locals are nearly identical and thus the method cannot 
be applied to resolve the degeneracy, either.

The third test is to compare the relative lens-source proper motions $\mu$ of the two degenerate 
solutions. If they differ by an amount substantially greater than the measurement error, it would 
be possible to resolve the degeneracy from future follow-up observation by using high-resolution 
space or ground-based instrument. We estimate the proper motions by $\mu=\theta_{\rm E}/t_{\rm E}$, 
where the Einstein time scale $t_{\rm E}$ is measured from light curve modeling and the angular 
Einstein radius $\theta_{\rm E}$ is estimated from the angular source radius $\theta_\ast$ and the normalized source 
radius $\rho_\ast$ by $\theta_{\rm E}=\theta_{\ast}/\rho_{\ast}$. The angular source radius is 
estimated based on the dereddened color and brightness of the source. The measured values are
$\mu=2.91\pm0.27$ $(0.16)$ mas yr$^{-1}$ for the high mass-ratio solution and $\mu=3.68\pm0.30$ 
$(0.15)$ mas yr$^{-1}$ for the low mass-ratio solution. We present two sets of errors where one (in the 
parenthesis) is estimated just based on the MCMC chain of the solution, while the other value is
estimated by adding additional 7\% error in quadrature to account for errors accompanied in the 
color-to-$\theta_\ast$ conversion process. The fractional error of the proper-motion difference 
is $\sigma_{\Delta\mu}/\Delta\mu\sim50\%$ $(28\%)$. Considering the large fractional error, it 
would not be easy to resolve the degeneracy by using this method.

Although very degenerate with the current data, however, we find that the latter degeneracy could have 
been resolved if the perturbation had been continuously and precisely covered by additional data. This can be seen 
in the model light curves of the two degenerate solutions presented in Figure~\ref{fig:one} (solid 
curve for the high mass-ratio solution and dotted curve for the low mass-ratio solution). It is found 
that the difference between the two model light curves in the region $6038.27\lesssim {\rm HJD}-2450000\lesssim 
6038.68$ is considerable, with a maximum magnitude difference reaching $\sim0.08$ magnitude. Although 
a portion of this region $6038.56\lesssim {\rm HJD}-2450000\lesssim 6038.64$ was covered by the KKO data, 
the event was still quite faint given the smaller 
aperture (36cm) of the telescope and thus the signal-to-noise ratio was not high enough to distinguish between models. 
Considering that photometric errors of adjacent data taken by 1-m class telescopes are $\sim 0.01$ 
magnitude, the degeneracy could have been easily resolved if the perturbation had been continuously 
covered by mid-size telescopes. Therefore, the degeneracy can be attributed to the incomplete coverage 
of the planetary perturbation. Considering this, the degeneracy is different from the case where degenerate 
light curves are alike in all parts.

\section{Conclusion}

We analyzed the high magnification microlensing event OGLE-2012-BLG-0455/MOA-2012-BLG-206 for 
which the peak of the light curve exhibited anomaly. Despite a large deviation from a 
standard point-mass model, it was found that there existed 4 very degenerate local solutions. 
While two of these were due to the well-known $s\leftrightarrow s^{-1}$ ``close/wide'' 
degeneracy, the other degeneracy, between high and low mass ratios q, was previously unknown. 
From the fact that the model light curves of the latter degeneracy substantially differed in 
the parts that were not covered by observation, it was found that the degeneracy was caused 
by the incomplete coverage of the perturbation. Therefore, the event illustrated the 
importance of continuous coverage of perturbations for accurate determinations of lens 
properties.

It is expected that the frequency of the degeneracy introduced in this work will be greatly 
reduced with the improvement of lensing surveys. Recently, there has been such improvement 
for the existing lensing surveys. For example, The observation cadence of the OGLE lensing 
survey was substantially increased with the adoption of a new wide field camera. The recent 
joint of the Wise survey \citep{shvartzvald14} being conducted in Israel enables more continuous 
event coverage by filling the gap between telescopes in Oceania and Chile. Furthermore, there 
are plans for future lensing surveys. For example, the MOA group plans to additionally locate 
a new telescope in Africa for better coverage of lensing events. In addition, the Korea Microlensing 
Telescope Network (KMTNet) will start operation from the 2014 season by using a network of telescopes 
at three different locations of the Southern Hemisphere (Chile, South Africa, and Australia). 
The KMTNet project plans to achieve 10-minute cadence. In addition to survey experiments, there 
also has been important progress in follow-up experiments. The most important is the completion 
of Las Cumbres Observatory Global Telescope Network, which is an integrated set of robotic telescopes 
distributed around the world, including two 2-m telescopes in Hawaii and Australia and nine 1-m 
telescopes sited in Chile, South Africa, Australia, and Texas \citep{tsapras09}. With the expansion 
of both survey and follow-up experiments, round-the-clock coverage of lensing events will be possible 
and the occurrence of the degeneracy will be greatly decreased.

\acknowledgments
Work by C.H. was supported by Creative Research Initiative Program
(2009-0081561) of National Research Foundation of Korea. A.G. and 
B.S.G. acknowledge support from NSF AST-1103471. B.S.G., A.G., and 
R.W.P. acknowledge support from NASA grant NNX12AB99G. The OGLE project has received 
funding from the European Research Council under the European Community's 
Seventh Framework Programme (FP7/2007-2013)/ERC grant agreement No. 
246678 to A.U. The MOA experiment was supported by grants JSPS22403003 
and JSPS23340064. T.S. acknowledges the support JSPS 24253004. T.S. 
is supported by the grant JSPS23340044. Y.M. acknowledges support 
from JSPS grants JSPS23540339 and JSPS19340058. S.D. was supported 
through a Ralph E. and Doris M. Hansmann Membership at the IAS and NSF 
grant AST-0807444. Work by J.C.Y. was performed in part under contract 
with the California Institute of Technology (Caltech) funded by NASA 
through the Sagan Fellowship Program.

\end{document}